\documentclass[aps,prd,groupedaddress, nofootinbib,showpacs, twocolumn]{revtex4}
\begin{document}
\title{$2+1$ gravity and Doubly Special Relativity }
\author{Laurent Freidel}
\email{lfreidel@perimeterinstitute.ca}
\affiliation{Perimeter Institute for Theoretical Physics \\
Waterloo Canada\\
ENS-Lyon, Lyon, France}
\author{Jerzy Kowalski-Glikman}
\email{jurekk@ift.uni.wroc.pl}
\thanks{Research  partially supported
by the    KBN grant 5PO3B05620.} \affiliation{Institute for
Theoretical Physics\\ University of Wroc\l{}aw, Poland}
\author{Lee Smolin}
\email{lsmolin@perimeterinstitute.ca}
\affiliation{Perimeter Institute for Theoretical Physics \\
Waterloo, Canada\\
Department of Physics, University of Waterloo, Waterloo, Canada}
\begin{abstract}
It is shown that gravity in 2+1 dimensions coupled to point
particles provides a nontrivial example of Doubly Special
Relativity (DSR). This result is obtained by interpretation of
previous results in the field and by exhibiting an explicit
transformation between the phase space algebra for one particle in
2+1 gravity found by Matschull and Welling and the corresponding
DSR algebra.  The identification of $2+1$ gravity as a $DSR$
system answers a number of questions concerning the latter, and
resolves the ambiguity of the basis of the algebra of observables.

Based on this observation a heuristic argument is made that the
algebra of symmetries of ultra high energy particle kinematics in
3+1 dimensions is described by some DSR theory.
\end{abstract}
 \maketitle

\section{Introduction}

Recently a proposal has been much discussed concerning how quantum
theories of gravity  may be tested experimentally. The {\it
doubly} or {\it deformed special relativity} proposal
($DSR$)\footnote{Aspects of $DSR$ theories have been proposed or
studied more than once
in the past, only to be forgotten and then rediscovered again.
Early formulations were by Snyder \cite{snyder} and Fock
\cite{fock}. During the 1990's the mathematical side of the
subject was developed under the name of $\kappa$-Poincar\'e
symmetry \cite{qP}, \cite{kappaP}. The recent interest is due to
the proposal that the effects of such theories may be both
testable and derivable from some versions of quantum gravity, see
for example \cite{Amelino-Camelia:2000ge} --
\cite{Magueijo:2001cr}.} is that quantum gravity effects may lead
in the limit of weak fields to modifications in the kinematics of
elementary particles characterized by
\cite{Amelino-Camelia:2000ge} -- \cite{Magueijo:2001cr}

\begin{enumerate}

\item{}Preservation of the relativity of inertial frames.

\item{}Non-linear modifications of the action of Lorentz boosts on
energy-momentum  vectors, preserving a preferred energy scale,
which is naturally taken to be the Planck energy, $E_p$.  In some
cases $E_p$ is a maximum mass and/or momentum that a single
elementary particle can attain.

\item{}Non-linear modifications of the energy-momentum relations, because
the function of $E$ and $\vec{p}$ that is preserved under the
exact action of the Lorentz group is no longer quadratic. This
could result in Planck scale effects such as an energy-dependent
speed of light and modifications of thresholds for scattering,
that may be observable in present and near future experiments.

\item{}Modifications in the commutators of coordinates and momentum and/or
non-commutativity of space-time coordinates.

\end{enumerate}

Theories with these characteristics are invariant under
modifications of the Poincar\'e algebra, called generically
$\kappa$-Poincar\'e algebras, where $\kappa$ is a dimensional
parameter that measures the deformations, usually taken to be
proportional to the Planck mass.

In a recent paper \cite{Amelino-Camelia:2003xp}, it was argued
that quantum gravity in $2+1$ dimensions \cite{Staruszkiewicz},
\cite{Deser:tn} with vanishing cosmological constant must be
invariant under some version of a $\kappa$-Poincar\'e symmetry.
The argument there depends only on the assumption that quantum
gravity in $2+1$ dimensions with the cosmological constant
$\Lambda =0$ must be derivable from the $\Lambda \rightarrow 0$
limit of $2+1$ quantum gravity with non-zero cosmological
constant. The argument is simple and algebraic, the point is that
 the symmetry which characterizes
quantum gravity in $2+1$ dimensions with $\Lambda >0$ is actually
quantum deformed de Sitter to $SO_{q}(3,1)$, with the quantum
deformation parameter $q$ given by \cite{Witten:1988hc},
\cite{Regge} \cite{Buffenoir:2002tx}
\begin{equation}
z = \ln (q) \approx l_{Planck}\sqrt{\Lambda}
\end{equation}
The limit $\Lambda \rightarrow 0$ then affects both the scaling of
the translation generators as the De Sitter group is contracted to
the Poincar\'e group, and the limit of $q\rightarrow 1$. It is
easy to see that because the ratio $\kappa=  \hbar
\sqrt{\Lambda}/z = G_{2+1}^{-1}$, where $G^{2+1}$ is Newton's
constant in $2+1$ dimensions, is held fixed, the limit gives the
$\kappa$-deformed symmetry group in $2+1$ dimensions. The
conclusion is that the symmetry algebra of $2+1$ dimensional
quantum gravity with $\Lambda =0$ is not Poincar\'e, it is a
$\kappa$-deformed Poincar\'e algebra. This means that the theory
must be a $DSR$ theory.

Quantum gravity in $2+1$ dimensions has been the subject of much
study in both the classical and quantum domain, beginning with the
work of Deser, Jackiw and 't Hooft \cite{Deser:tn} --
\cite{Matschull:2001ec}.  If that theory is a $DSR$ theory than
the features just listed above must be present, and this could not
have been easily missed by investigators.

Indeed, {\it all of the listed features have been seen in the
literature on $2+1$ gravity.}  In the next section we review some
of the long standing results in $2+1$ gravity and show how they
may be understood using the langauge of $DSR$ theories. To clinch
the relation, in section 3 we exhibit an explicit mapping between
the phase space  of  quantum gravity in $2+1$ dimensions  coupled
to a single point particle, studied in \cite{Matschull:1997du},
and the algebra of symmetry generators of a $DSR$ theory.

The observation that $2+1$ gravity provides examples of $DSR$
theories can help the study of both sides of the relation. The
language of $DSR$ theories and their foundations in terms of
general principles can unify and explain some results in the
literature of $2+1$ dimensional gravity that, when first
discovered, seemed strange and unintuitive. We can now see that
some of the features of $2+1$ gravity are neither strange nor
necessarily unique to $2+1$ dimensions, because they follow only
from the general requirement that the transformations between
different inertial frames preserve an energy scale.

Furthermore, what one has in the $2+1$ gravity models, such as
those with gravity coupled to $N$ point particles, is a class of
non-trivial $DSR$ theories that are completely explicit and
solvable, both classically and quantum mechanically. The existence
of these examples answers a number of questions and challenges
that have been raised concerning $DSR$ theories.  Some authors
have argued \cite{ahluwalia} that $DSR$ theories are just ordinary
special relativistic theories rewritten in terms of some
non-linear combinations of energy and momentum, while, conversely,
others have argued that they must be trivial because interactions
cannot be consistently included. Both criticisms are shown wrong
by the existence of an explicit and solvable class of $DSR$
theories, with interactions, given by quantum gravity in $2+1$
dimensions coupled to point particles and fields.

Furthermore, we see that in $2+1$ dimensions the apparent problem
of the freedom to choose the basis of the symmetry algebra of a
$DSR$ theory is resolved by the fact that the choice of the
coupling of matter to the gravitational field picks out the
physical energy and momentum.  We see in section 3 below that for
the case of minimal coupling of gravity to a single point particle
the basis picked out  is the classical basis.

Finally, one can ask whether the fact that $2+1$ gravity is
a $DSR$ theory has any implications for real physics in $3=1$
dimensions. In the final section of the paper we present a
heuristic argument that it may.

\section{Signs of $DSR$ in $2+1$ gravity}

In this section we point out where effects characteristic of $DSR$
have been discovered already in the literature on $2+1$
dimensional gravity.  We consider only the case $\Lambda=0$.

\begin{itemize}

\item{}It is important first to note that
Newton's constant in $2+1$ dimensions, denoted here by $G$, has
dimensions of inverse mass (with only $c=1$ and no $\hbar$
involved)\footnote{$G$ is identified with inverse of the $\kappa$
deformation parameter of $\kappa$-Poincar\'e algebra.}. Thus, if
the asymptotic symmetry group knows about gravity, it will have to
preserve the scale $G^{-1}$. Of course, in theories with
sufficiently short range interactions the asymptotic symmetry
group does not depend on the coupling constants. But in $2+1$
gravity the presence of matter causes the geometry of spacetime to
become conical and this deforms the asymptotic conditions in a way
that depends on $G$.  Further, since $\hbar$ is not involved in
the definition of the mass scale, $G^{-1}$, the deformation
affects also the algebra of the classical phase space. This is the
main reason why $2+1$ gravity is a $DSR$ theory.

\item{}In $2+1$ gravity coupled to point particles, the hamiltonian, $H$,
whose value is equal to the $ADM$ mass, and hence is measured by a
surface term, is bounded from both above and below
\cite{Matschull:1997du}, \cite{Matschull:2001ec},
\begin{equation}\label{s1} 0 \leq H \leq {1\over
4G}
\end{equation}
This can be understood in the following way. In $2+1$ dimensions
the spacetime is flat, except where matter is present. A particle,
or in fact any compactly supported distribution of matter, is
surrounded by an asymptotic region, which is locally flat, and
whose geometry is thus characterized by a deficit angle $\alpha$.
A standard result is that \cite{Deser:tn},
 \cite{t Hooft}, \cite{Buffenoir:2002tx},
 \cite{Matschull:1997du}, \cite{Matschull:2001ec}, \cite{Ashtekar:1993ds}
\begin{equation}
\alpha = 8 \pi G H
\end{equation}
But a deficit angle $\alpha$ must be less than or equal to $2\pi$.
Hence there is an upper limit on the mass of any system, as
measured by the hamiltonian. The upper limit holds for all
systems, regardless of how many particles there are and what their
relative positions or motions are.

This upper mass limit must be preserved by the asymptotic symmetry
group. Hence the asymptotic symmetry group cannot be the ordinary
Poincar\'e group, it must be a $DSR$ theory with a maximum energy.

\item{}It has further been shown that the spatial components of momentum of a
particle in $2+1$ gravity are unbounded \cite{Matschull:1997du}.
This, together, with a bounded energy, implies a modified
energy-momentum relation.

\item{}The phase space of a single point particle in $2+1$ gravity was
constructed by Matschull and Welling in \cite{Matschull:1997du}
and it was found that a classical solution is labelled by a three
dimensional position ${\cal Y}_\mu$, $\mu=0,1,2$ and momentum
$p_\mu$. They find explicitly that the energy momentum relations
and the action of the Poincar\'e symmetry are deformed, in a way
that preserves a fixed energy scale. They indeed make explicit
reference to the work of Snyder \cite{snyder}, which was an early
proposal for $DSR$.

\item{}Furthermore, Matschull and Welling find that
the spacetime  coordinates ${\cal Y}_\mu$ of a particle are
non-commutative under the classical Poisson brackets,
\begin{equation}
 [{\cal Y}_\mu, {\cal Y}_\nu] = -{2}G\,\epsilon_{\mu\nu\rho}\,{\cal
 Y}_\rho.
\end{equation}
This property was found in \cite{Matschull:2001ec} to extend to
systems of $N$ particles.

\item{}Matschull and Welling also find that the
components of the energy-momentum vector for a point particle in
$2+1$ gravity live on a curved manifold, which is $2+1$
dimensional Anti-de Sitter spacetime.  This was shown in
\cite{Kowalski-Glikman:2002jr}, \cite{Kowalski-Glikman:2003we} to
be a feature of $DSR$ theories\footnote{Although in reference
\cite{Kowalski-Glikman:2002jr}, \cite{Kowalski-Glikman:2003we} the
momentum space for a class of $DSR$ theories was shown to be de
Sitter spacetime. We discuss below the difference between
positively and negatively curved momentum spaces.}

\item{}Ashtekar and Varadarajan \cite{Ashtekar:1993ds} found a relationship
between two
definitions of energy relevant for $2+1$ gravity, which is
reminiscent of non-linear redefinitions of the energy used in
changing bases between different realizations of $DSR$ theories.
The case they studied has to do with $3+1$ gravity, with two
Killing fields, one rotational and one axial. One first
dimensionally reduces to $2+1$ dimensions, in which case the
dynamics of GR in $3+1$ is expressed as a scalar field coupled to
$2+1$ dimensional $GR$. The ADM Hamiltonian $H$ still exists and
still is bounded from above as in (\ref{s1}). But in the presence
of the additional, rotational Killing field, the theory can be
represented by a scalar field evolving in a flat reference
Minkowski spacetime, with the ordinary hamiltonian
\begin{equation}
H_{flat}= {1 \over 2} \int_0^\infty dr r \left [ \dot{\phi}^2 +
(\partial_r \phi )^2 \right ]
\end{equation}

$H_{flat}$ is of course unbounded above. They find the
relationship between them is
\begin{equation}\label{s2}
H= {1\over 4G} \left ( 1-e^{-4GH_{flat}} \right )
\end{equation}

This exact relation is in fact present in the literature on $DSR$
\cite{Kowalski-Glikman:2002jr}. It holds in the a presentation of
the $\kappa$-Poincar\'e algebra known as the ``bicrossproduct''
basis. In that case $H_{flat}=E$ is, as in the present case, the
zeroth component of an energy momentum vector and $H$ is the
``physical rest mass,$m_{0}$ defined by,
\begin{equation} {1\over m_{0}}= \lim_{p \rightarrow
0}\left. {1\over p}{dE \over dp}\right|_{E=p_{0}}
\end{equation}

It is intriguing that this is  the inertial mass, while, for the
solutions with rotational symmetry, the $ADM$ energy is the active
gravitational mass. Since they are both expressed in terms of the
zeroth component of the energy-momentum vector by the same
equation, they coincide on the subset of solutions on which they
are both defined, which are the rotationally invariant solutions.
This appears to be a direct demonstration of the equality of
inertial and gravitational mass, within this context.

Indeed, this observation suggests that the Ashtekar-Varadarajan
form of the $ADM$ mass is more general than their calculation
shows. Indeed it is not hard to see that this is the case.  Let us
study the free scalar field in $2+1$ dimensional Minkowski
spacetime, with no condition of rotational symmetry. This system
is {\it not} a dimensional reduction of general relativity, only a
subspace of solutions, those with rotational symmetry, are related
to general relativity.  But it still may serve as a useful example
of a DSR theory. Of course the theory has full Poincar\'e
invariance, with momentum generators $P_{i}$ and boost generators
$K_{i}$ satisfying the usual Poincar\'e algebra. But eq.
(\ref{s2}) implies that they form with $H$ a $DSR$ algebra
\begin{eqnarray}
\{ K_{i},H \}  &=& (1-4GH)P_{i}
\nonumber \\
\{ K_{i}, P_{j}\} &=& -{1 \over 4G} \delta_{ij} \ln [1-4GH]
\end{eqnarray}
with the other commutators undeformed. The physical energy
momentum relations are deformed to
\begin{equation}
P_{i}^{2}+m^{2} = {1\over 16G^{2}} [\ln (1-4GH)]^{2}
\end{equation}

\item{} Recent calculations \cite{Bais:2002ye} indicate that
quantum deformations of symmetries play a role in gravitational
scattering of particles in 2+1 dimensions.

\end{itemize}

All of these pieces of evidence show that $2+1$  gravity coupled
to matter can be understood as a $DSR$ system.

Of course, the $2+1$ dimensional model system is not completely
analogous to real physics in $3+1$ dimensions. But this result
answers cleanly several queries and  criticisms that have been
levied against the $DSR$ proposal.

First, some authors have suggested that $DSR$  theories are
physically indistinguishable from ordinary special relativity
\cite{ahluwalia}. They argue that in some cases, one can arrive at
a $DSR$ system from a non-linear mapping of energy-momentum space
to itself. These results show that argument fails, for  there is
no doubt that the model system of point particles in $2+1$ gravity
is physically distinguishable from the model system of free
particles in flat $2+1$ dimensional spacetime. This is here a
clean result, with no quantization ambiguities, because the
deformation parameter $\kappa=1/4G$ is entirely classical and the
modification is of the structure of the classical phase space. The
two phase spaces are not isomorphic, when gravity is turned on,
the phase space is curved, but when $G=0$ the phase space is flat.

This is clear also for the multiparticle system,  where there are
non-trivial interactions, depending on $G$, which make the system
measurably distinct from the free particle case with $G=0$.

The multiparticle system in $2+1$ gravity also serves as  an
example of a counterintuitive property of some $DSR$ models in
$3+1$ gravity. This is that the upper mass limit $M_{upper} =1/4G$
is independent of the number of particles in the system. This of
course cannot be the case in the real world, so it is good to know
that there are implementations of $DSR$ in $3+1$ dimensions that
do not have an upper mass limit for systems of many particles, or
where the upper mass limit grows with the number of particles or
the mass of the total system, in such a way as to not violate
experience \cite{Magueijo:2002am}.

However, it is also good to know that  there is a model system,
which is sensible physically, in which this non-intuitive feature
is completely realized. Moreover it suggests the start of a
physical answer to one of the puzzling questions about $DSR$
models. This is that the addition of energy and momentum in $DSR$
theories is non-linear. This can be understood as a consequence of
the non-linear action of the Lorentz group, for example it follows
from the fact that the energy-momentum space has non-zero
curvature. It appears to remain even in realizations of $DSR$ that
remove the mass limit for composite systems.

Some physicists have criticized the $DSR$  proposal by pointing
out that the non-linear corrections to addition of energy-momentum
vectors for a system of two particles can be interpreted by saying
that there is a binding energy between pairs of particles that
does not depend on the distance between them, but depends only on
the individual energies and momenta.

This may be counter-intuitive, but it  is precisely the what
happens in $2+1$ dimensions. Because spacetime is locally flat,
each particle contributes a deficit angle to the overall geometry
that affects all the other particles' motions, no matter how far
away. The result is that there is a binding energy that is
independent of distance.

This suggests a speculative remark:  might there be even in $3+1$
dimensions a small component of the binding energy of pairs of
particles, of order $l_p M_1M_2$, which is independent of
distance? Might this be interpreted as a kind of quantum gravity
effect?

In the last section we make some  speculative remarks concerning
the question of whether these results have any bearing on real
physics in $3+1$ dimensions.

\section{Phase space of DSR in 2+1 dimensions}

In this section we will compare the phase space of 2+1 dimensional
DSR with that of 2+1 dimensional gravity with one particle. Let us
start with the former.

\subsection{Phase spaces of DSR}

As in $3+1$ dimensions,  the starting point to find the phase
space of  DSR theory in the 2+1 dimensional case is the $2+1$
dimensional $\kappa$-Poincar\'e algebra \cite{kappaP}, the quantum
algebra whose generators are momenta\footnote{The Greek indices
run from 1 to 3,  the Latin ones from 1 to 2, while the capital
ones from 0 to 3.} $p_\mu=(p_0, p_i)$ and Lorentz algebra
generators $J_\mu= (M, N_i)$ boosts. Taking the co-algebra of the
$\kappa$-Poincar\'e quantum algebra and using the so-called
``Heisenberg-double construction'' \cite{crossalg}, \cite{luno},
\cite{Kowalski-Glikman:2002jr} it is possible to derive the
position variables, conjugate to momenta, $x_\mu$, as well as the
brackets between them and the $\kappa$-Poincar\'e algebra
generators.

This (quantum) algebraic construction has a geometrical
counterpart, described in \cite{Kowalski-Glikman:2002ft},
\cite{Kowalski-Glikman:2003we}. Here the manifold on which
momenta live is de Sitter space (in the case at hand the 3
dimensional one). The positions and the Lorentz transformations
are symmetries acting on the space of momenta. Thus they form the
three dimensional de Sitter algebra $SO(3,1)$. It is convenient to
define the de Sitter space of momenta as a three dimensional
surface
\begin{equation}\label{1}
 -\eta_0^2 + \eta_1^2 + \eta_2^2 + \eta_3^2 = \kappa^2
\end{equation}
in the four dimensional Minkowski space with coordinates
$(\eta_0,\ldots \eta_3)$. The physical momenta $p_\mu$ are then
the coordinates on the surface (\ref{1}). This means that we can
think of $\eta_A= \eta_A(p_\mu)$ as of the given functions of
momenta, for which the equation (\ref{1}) is identically
satisfied. In the DSR terminology, the choice of a particular
coordinate system on de Sitter space corresponds to a choice of
the so called DSR basis (see \cite{Kowalski-Glikman:2002jr},
\cite{Kowalski-Glikman:2003we}). It turns out that in order to
relate DSR to the 2+1 gravity one has to choose the so called
classical basis, characterized by $\eta_\mu = p_\mu$. This choice
will be implicit below, however we find it more convenient to
write down the formulas below in terms of the variables $\eta_A$.

The algebra of symmetries of the de Sitter space of momenta
(\ref{1}) can be most easily read off by writing down the action
of these symmetries on the four-dimensional Minkowski space with
coordinates $\eta_A$ and then pulling them down to the surface
(\ref{1}). Let us note however that while it is easy to identify
the Lorentz generators $J_\mu=(M, N_i)$ as the elements of the
$SO(2,1)$ subalgebra of the $SO(3,1)$, it is a matter of
convenience which linearly independent combination of generators
is to be identified with positions (i.e.~the generators of
translation in momentum space.) Technically speaking we are free
to choose the decomposition of $SO(3,1)$ into the sum of $SO(2,1)$
and its remainder.

In the case of the DSR phase space, the action of the symmetries
is given by
\begin{equation}\label{2}
  [M, \eta_i] = \epsilon_{ij}\eta_j, \quad [N_i, \eta_j] =
  \delta_{ij}\, \eta_0, \quad [N_i, \eta_0] =   \eta_i,
\end{equation}
\begin{equation}\label{3}
[J_\mu,\eta_3] =0,
\end{equation}
with $J_\mu$ satisfying the algebra
\begin{equation}\label{xx}
 [M, N_i] =\epsilon_{ij}\, N_j, \quad [N_i,N_j] = -
 \epsilon_{ij}\, M
\end{equation}

\begin{equation}\label{4}
  [x_0,\eta_3] = \frac{1}\kappa\, \eta_0, \quad [x_0,\eta_0]
  = \frac{1}\kappa\, \eta_3, \quad [x_0,\eta_i] = 0,
\end{equation}
\begin{equation}\label{5}
  [x_i, \eta_3] = [x_i, \eta_0] =\frac{1}\kappa\, \eta_i,
  \quad [x_i, \eta_j] = \frac{1}\kappa\,
\delta_{ij}(\eta_0 - \eta_3),
\end{equation}

Note that it follows from these equations that
\begin{equation}\label{6}
  [x_0, x_i] = -\frac{1}{\kappa}\, x_i, \quad [x_i, x_j] = 0.
\end{equation}
It is worth mentioning also  that such a  decomposition is
possible in any dimension. In particular in the 3+1 case the
bracket (\ref{6}) descries the so-called $\kappa$-Minkowski type
of non-commutativity.

One can repeat this geometric construction in the case when the
momenta manifold is the anti de Sitter space
\begin{equation}\label{1a}
 -\eta_0^2 + \eta_1^2 + \eta_2^2 - \eta_3^2 = \kappa^2.
\end{equation}
Now the symmetry algebra is $SO(2,2)$, having again the three
dimensional Lorentz algebra $SO(2,1)$ described by (\ref{2}),
(\ref{3}) as its subalgebra. The algebra of positions, which we
denote $y_\mu$ (i.e. translations of momenta) changes only
slightly and now reads
\begin{equation}\label{4a}
  [y_0,\eta_3] = -\frac{1}\kappa\, \eta_0, \quad [y_0,\eta_0] =
  \frac{1}\kappa\, \eta_3, \quad [y_0,\eta_i] = 0,
\end{equation}
$$
  [y_i, \eta_3] =- \frac{1}\kappa\, \eta_i, \quad [y_i, \eta_0] =
  \frac{1}\kappa\, \eta_i,
$$
\begin{equation}\label{5a}
[y_i, \eta_j] = \frac{1}\kappa\, \delta_{ij}(\eta_0 - \eta_3),
\end{equation}
From (\ref{4a}), (\ref{5a}) it  follows that
\begin{equation}\label{6a}
[y_0, y_i] = - \frac1\kappa\, y_i + \frac1{\kappa^2}\, N_i, \quad
[y_i, y_j] = - \frac2{\kappa^2}\, \epsilon_{ij}\, M.
\end{equation}
We see that the bracket (\ref{6a}) does not describe the
$\kappa$-Minkowski type of non-commutativity. Since the
non-commutativity type is related to the co-algebra structure of
the quantum Poincar\'e algebra, this result indicates that along
with the $\kappa$-Poincar\'e algebra there exists another quantum
Poincar\'e algebra with the same algebra, but different
co-algebra, which we expect to be related to the former by a
twist\footnote{This expectation is based on the classification of
Poisson structures on Poincar\'e group presented in
\cite{zakrzewski1997}.}

\subsection{Phase space of 2+1 gravity}

The phase space algebra of one particle in  2+1 dimensional
gravity is the algebra of asymptotic charges. This algebra has
been carefully analyzed by Matschull and Welling in
\cite{Matschull:1997du}. They find that the physical momentum
manifold is  anti de Sitter space and that $\eta_\mu = p_\mu$, as
stated above. This means that 2+1 gravity seems to pick the
classical basis of DSR as the one having physical relevance.
Further, Matschull and Welling employ a particular decomposition
of the $SO(3,1)$ algebra, in which  the positions ${\cal Y}_\mu$
act on momenta as right multiplication and have the following
brackets with $\eta_A$.
\begin{equation}\label{7}
 [{\cal Y}_0, \eta_3] =  -\frac{1}{\kappa}\, \eta_0, \quad [{\cal Y}_0, \eta_0] =
 \frac{1}{\kappa}\, \eta_3, \quad [{\cal Y}_0, \eta_i] = -\frac{1}{\kappa}\, \epsilon_{ij}\, \eta_j
\end{equation}
$$
[{\cal Y}_i, \eta_3] =  -\frac{1}{\kappa}\, \eta_i, \quad [{\cal
Y}_i, \eta_0] = \frac{1}{\kappa}\, \epsilon_{ij}\, \eta_j ,
$$
\begin{equation}\label{8} [{\cal Y}_i, \eta_j] =
\frac{1}{\kappa}\left(\epsilon_{ij}\, \eta_0 -\delta_{ij}\,
\eta_3\right) .
\end{equation}

Comparing the expressions (\ref{4a}), (\ref{5a}) with (\ref{7}),
(\ref{8}) we easily find that these decompositions are related by
\begin{equation}\label{9}
{\cal Y}_0 = y_0 -\frac{1}{\kappa}\, M,\quad {\cal Y}_i = y_i -
\frac1\kappa\left( N_i -\epsilon_{ij}\, N_j\right).
\end{equation}
It can be also easily checked that
\begin{equation}\label{10}
 [{\cal Y}_\mu, {\cal Y}_\nu] = -\frac{2}{\kappa}\,\epsilon_{\mu\nu\rho}\,{\cal
 Y}^\rho.
\end{equation}
Thus the DSR anti de Sitter phase space is (up to a trivial
reshuffling of the generators) equivalent to the phase space of a
single particle in 2+1 gravity.

It is an open problem whether one can get  de Sitter space as a
manifold of momenta from 2+1 quantum gravity. It would be
interesting to see if this is the case. If so, there exist two
kinds of phase spaces of a particle in a 2+1 gravitational field
corresponding to two DSR phase space algebras presented above.

\section{Implications for physics in $3+1$ dimensions}

We present here an argument that suggests  that the results of
this paper, and of those we reference, concerning $2+1$
dimensional quantum gravity coupled to point particles may have
implications for real physics in $3+1$ dimensions.

The main idea is to construct an experimental  situation that
forces a dimensional reduction to the $2+1$ dimensional theory. It
is interesting that this can be done in quantum theory, using the
uncertainty principle as an essential element of the argument.

Let us consider a system of two relativistic  interacting
elementary particles in $3+1$ dimensions, whose masses are less
than $G^{-1}$.  In the center of mass frame the motion will be
planar.  Let us consider the system as described by an inertial
observer who travels perpendicular to the plane of the system's
motion, which we will call the $z$ direction.  From the point of
view of that observer, the system is in an eigenstate of total
longitudinal momentum, $\hat{P}^{total}_z$, with some eigenvalue
$P_{z}$. Since the system is in an eigenstate of
$\hat{P}^{total}_z$ the wavefunction of the center of mass will be
uniform in $z$.

Further, since there was initially zero relative momentum
between the particles in the $z$ direction it is also true
in the observers frame that
\begin{equation}
{P}^{rel}_z= {P}^{1}_z-{P}^{2}_z =0.
\end{equation}
This implies of course ${P}^{1}_z= {P}^{2}_z = {P}^{total}_z/2$.
Then the above applies as well to each particle, i.e. their
wavefunctions are uniform in the $\hat{z}$ direction as their
wavefunctions have wavelength $2L$ where
\begin{equation}
L= {\hbar \over P_{z}^{total} }
\end{equation}

At the same time, we assume that the  uncertainties in the
transverse positions are bounded a scale $r$,
such that $ r \ll 2L $.

Then the wavefunctions for the two particles have support on
narrow cylinders of radius $r$ which extend
uniformly in the $z$ direction.

Finally, we assume that the state of the  gravitational field is
semiclassical, so that to a good approximation, within $\cal C$
the semiclassical Einstein equations hold.
\begin{equation}
G_{ab}= 8\pi G <\hat{T}_{ab} >
\label{semi}
\end{equation}

Note that we do not have to assume that the  semiclassical
approximation holds for all states. We assume something much
weaker, which is that there are subspaces of states in which it
holds.

Since the wavefunctions are uniform in $z$, this implies that
the gravitational field seen by our observer will have a spacelike Killing
field $k^a= (\partial /\partial z)^a$.

Thus, if there are no forces other than the gravitational field, the
scattering of the two particles described semiclassically by
(\ref{semi}) must be the same as that of two parallel cosmic strings.
This is known to be described by an equivalent $2+1$ dimensional
problem in which the gravitational field is dimensionally reduced
along the $z$ direction so that the two ``cosmic strings'' which
are the sources of the gravitational field, are replaced by two
punctures.

The dimensional reduction is governed by a length $d$, which is
the extent in $z$ that the system extends. We cannot take $d<L$
without violating the uncertainty principle. It is then convenient
to take $d=L$.  Further, since the system consists of elementary
particles, they have no intrinsic extent, so there is no other
scale associated with their extent in the $z$ direction. We can
then identify $z=0$ and $z=L$ to make an equivalent toroidal
system, and then dimensionally reduce along $z$. The relationship
between the four dimensional Newton's constant $G^{3+1}$ and the
three dimensional Newton's constant $G^{2+1}=G$, which  played a
role so far in this paper is given by
\begin{equation}
G^{2+1} = {G^{3+1} \over 2L} = {G^{3+1} P^{tot}_z \over 2\hbar}
\end{equation}

Thus, in the analogous $2+1$ dimensional system, which is equivalent
to the original system as seen
from the point of view of the boosted observer, the Newton's
constant depends on the longitudinal momenta.

Of course, in general there
will be an additional scalar field, corresponding to the
dynamical degrees of freedom of the gravitational field. We will
for the moment assume that these are unexcited, but exciting them
will not affect the analysis so long as the gravitational
excitations are invariant also under the killing field and are of
compact support.

Now we note that, if there are no other particles or excited
degrees of freedom, the energy of the system  can
to a good approximation be described by the hamiltonian $H$ of the
two dimensional dimensionally reduced system. This is described
by a boundary integral, which may be taken over any circle that
encloses the two particles.
But this is bounded
from above, by (\ref{s1}). This may seem strange, but it is easy
to see that it has a natural four dimensional interpretation.

The bound is given by
\begin{equation}
M < {1 \over 4 G^{2+1} } = {2L \over 4 G^{3+1} }
\end{equation}
where $M$ is the value of the ADM hamiltonian, $H$. But this just
implies that
\begin{equation}
L > 2G^{3+1}M = R_{Sch} \label{sch}
\end{equation}
i.e. this has to be true, otherwise the dynamics of the
gravitational field in $3+1$ dimensions would have collapsed the
system to a black hole!  Thus, we see that the total bound from
above of the energy in $2+1$ dimensions is necessary so that one
cannot violate the condition in $3+1$ dimensions that a system be
larger than its Schwarzschild radius.

Note that we also must have
\begin{equation}
M > P^{tot}_z ={
\hbar \over L}
\end{equation}
Together with (\ref{sch}) this implies $L>
l_{Planck}$, which is of course necessary if the semiclassical
argument we are giving is to hold.

Now, we have put no restriction on any components of  momentum or
position in the transverse directions.  So the system still has
symmetries in the transverse directions.  Furthermore, the argument
extends to any number of particles, so long as their relative
momenta are coplanar. Thus, we learn the following.

Let ${\cal H}^{QG}$ be the full Hilbert space of the quantum
theory of gravity, coupled to some appropriate matter fields, with
$\Lambda=0$. Let us consider a subspace of states ${\cal
H}^{weak}$ which are relevant in the low energy limit in which all
energies are small in Planck units.  We expect that this will have
a symmetry algebra which is related to the Poincar\'e algebra
${\cal P}^{3+1}$ in $3+1$ dimensions, by some possible small
deformations parameterized by $G^{3+1}$ and $\hbar$. Let us call
this low energy symmetry group ${\cal P}^{3+1}_{G}$.

Let us now   consider the subspace of ${\cal H}^{weak}$ which is
described by the system we have just constructed . It contains two
particles, and is an eigenstate of $\hat{P}^{tot}_z$ with large
$P^{tot}_z$ and vanishing relative longitudinal momenta.
Let us call this subspace of Hilbert space
${\cal H}_{P_z}$.

The conditions that define this subspace break the  generators of
the (possibly modified) Poincar\'e algebra that involve the $z$
direction. But they leave unbroken the symmetry in the $2+1$
dimensional transverse space. Thus, a subgroup of ${\cal
P}^{3+1}_{G}$ acts on this space, which we will call ${\cal
P}^{2+1}_{G} \subset {\cal P}^{3+1}_{G}$.

We have argued that the physics in ${\cal H}_{P_z}$ is to good
approximation described by an analogue system in of two particles
in $2+1$ gravity. However, we know from the results cited in the
previous sections that the symmetry algebra acting there is not by
the ordinary $2+1$ dimensional Poincar\'e algebra, but by the
$\kappa$-Poincar\'e algebra in $2+1$ dimensions, with
\begin{equation}
\kappa^{-1}  = {4 G^{3+1} P^{tot}_z \over \hbar}
\end{equation}
In particular, there is a maximum energy given by
\begin{equation}
M_{max}(P^{tot}_z) = \kappa= { M_{Planck}^2 \over 4P^{total}_z}
\end{equation}
This gives us a last condition, \begin{equation} MP^{total}_z < {
M_{Planck}^2 \over 4} \end{equation} which is compatible with the
previous conditions. Thus, when all the conditions are satisfied,
the deformed symmetry algebra must be identified with ${\cal
P}^{2+1}_{G}$.

Now we can note the following. Whatever ${\cal P}^{3+1}_{G}$ is,
it must have the following properties:

\begin{itemize}
\item{} It depends on $G^{3+1}$ and $\hbar$, so that it's  action on
{\it each} subspace ${\cal H}_{P_z}$, for each choice of $P_z$, is
the $\kappa$ deformed $2+1$ Poincar\'e algebra, with $\kappa$ as
above.

\item{} It does not satisfy the rule that momenta and energy add, on
all states in $\cal H$, since they are not satisfied in these
subspaces.

\item{} Therefore, whatever $ {\cal P}^{3+1}_{G}$ is, it is not the
classical Poincar\'e group.
\end{itemize}

Thus the theory of particle kinematics at ultra high energies is
not Special Relativity, and the arguments presented above suggest
that it might be Doubly Special Relativity.

\section{Acknowledgement}

We would like to thank Giovanni Amelino-Camelia and Artem
Staroduptsev for discussions.

\end{document}